\documentclass[
  aps,reprint,
  notitlepage,
  amsfonts,
  amssymb,
  amsmath,
  superscriptaddress,
  floatfix,
  longbibliography
]{revtex4-1}
\usepackage{graphicx}
\usepackage[
  colorlinks,
  pdftitle={Polarization and plasmons in hot photoexcited graphene},
  pdfauthor={A. Freddie Page, Joachim M. Hamm, Ortwin Hess}
]{hyperref}
\hypersetup{
  linkcolor=[rgb]{0,0,1},
  citecolor=[rgb]{0,0,1},
   urlcolor=[rgb]{0,0,1}
}

\newcommand{\qv}{\mathbf{q}}
\newcommand{\kv}{\mathbf{k}}
\newcommand{\eps}{\varepsilon}
\newcommand{\epsrpa}{\eps}
\newcommand{\epsbar}{\bar{\eps}}
\newcommand{\eeps}{\epsilon} 
\newcommand{\vf}{v_\mathrm{F}}

\newcommand{\dd}{\mathrm{d}}

\newcommand{\pd}[2]{\frac{\partial #1}{\partial #2}}
\DeclareMathOperator{\res}{Res}
\DeclareMathOperator{\im}{Im}
\DeclareMathOperator{\re}{Re}
\DeclareMathOperator{\sech}{sech}
\newcommand{\sigs}{\sigma_\mathrm{s}}
\newcommand{\elec}{\mathrm{e}}
\newcommand{\hole}{\mathrm{h}}
\newcommand{\nE}{n_\elec}
\newcommand{\nH}{n_\hole}
\newcommand{\nEH}{n_{\elec/\hole}}
\newcommand{\muE}{\mu_\elec}
\newcommand{\muH}{\mu_\hole}
\newcommand{\omScale}{\omega_\mathrm{D}}

\newcommand{\epsScale}{\hbar\omScale}

\newcommand{\eV}{\mathrm{eV}}

\newcommand{\nSo}{n_{\mathrm{s}0}}
\newcommand{\epsS}{\eeps_\mathrm{s}}
\newcommand{\gamS}{\gamma_\mathrm{s}}

\newcommand{\omPl}{\omega_\mathrm{pl}} 
\newcommand{\gamPl}{\gamma_\mathrm{pl}} 

\newcommand{\alphag}{\alpha_\mathrm{g}}
\newcommand{\omtil}{\tilde\omega}
\newcommand{\qtil}{\tilde{q}}

\newcommand{\branch}{{\mathord{+\mkern-14mu\times}}}
\newcommand{\pole}{{\!\mathord{\times}\!}}

\newcommand{\imperial}{
Blackett Laboratory,
Department of Physics,
Imperial College London,
London SW7~2AZ,
United Kingdom
}

\begin{document}
\title{Polarization and plasmons in hot photoexcited graphene}
\author{A.~Freddie~Page}
\affiliation{\imperial}
\author{Joachim~M.~Hamm}
\affiliation{\imperial}
\author{Ortwin~Hess}
\email{o.hess@imperial.ac.uk}
\affiliation{\imperial}
\date{\today}


\begin{abstract}
We present a robust and exact method for calculating the polarization function
and plasmon dispersion of graphene, for an arbitrary (isotropic)
non-equilibrium carrier distribution, within random phase approximation (RPA).
This is demonstrated for a range of carrier distributions, including hot carrier
distributions which occur within the femtoseconds following photoexcitation.
We show that qualitatively different behaviour from the equilibrium case can
occur.
As the polarization function determines dynamic screening, its calculation shall
be essential to quantifying carrier-carrier scattering channels for graphene far
from equilibrium.
\end{abstract}

\maketitle

\begin{figure}
 \includegraphics{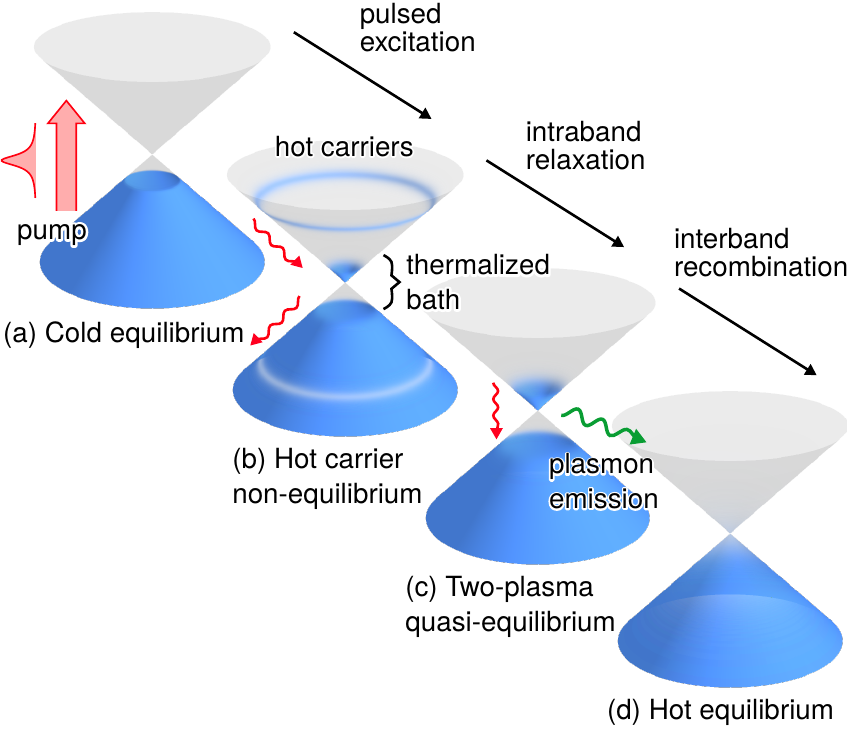}
 \caption{\label{concept}
 Highly energetic electron-hole pairs (hot carriers) are excited in graphene by
 a pulsed femtosecond beam.
 These carriers then relax by intraband processes on a 10-100fs timescale and
 pool at the Dirac point approaching a thermalized two-component
 quasi-equilibrium distribution, before relaxing to a warm carrier equilibrium
 where the electrons are in thermal equilibrium, though at a higher temperature
 than the phonons and ambient surroundings.
 Such varying electronic configurations determine the energy scale and character
 of the dynamic screening and the supported plasmon modes on the sheet, as well
 as whether net stimulated emission of plasmons is possible.
 }
\end{figure}

\section{Introduction}
Graphene stands out amongst other two-dimensional materials due to the presence
of Dirac points in its band structure, with an approximately linear electronic
dispersion and a vanishing band-gap.
One of the most important consequences of this band structure is the possibility
for low energetic electron/hole pairs to interact with plasmons via plasmon
absorption or emission;
This is excluded for in band-gap materials where the energy of plasmons is
typically much smaller than that of single particle excitations
\cite{Giuliani1982,Bostwick2007}.

When graphene is excited with a femtosecond optical pulse, high-energetic
electron/hole pairs are generated far from the Fermi edge.
Carrier-carrier and carrier-plasmon scattering will cause a rapid redistribution
of the energy in the carrier system, on a 10-100fs scale \cite{Song2015},
firstly bringing the plasma into a two-component inverted quasi-equilibrium  and
then into a state of a high-temperature carrier equilibrium
\cite{Dawlaty2008,Li2012,Zhang2013,Meng2016}.
Following the equilibration in the carrier system, the plasma will equilibrate
with the phonon system, albeit on longer timescales, as carrier-phonon
scattering is typically at least one order of magnitude slower than
carrier-carrier scattering \cite{Butscher2007,Rana2009,Wang2010}.
Excitation and relaxation processes are illustrated in Fig.~\ref{concept}.
During this relaxation process, the scattering of carriers and their interaction
with plasmons are mediated by the screened Coulomb potential
$V_\qv^\mathrm{eff}(q,\omega) = V_q /\eps(q,\omega)$, where $V_q $ is the
bare Coulomb potential and $\eps(q,\omega)$ the dynamic dielectric
function, which in turn depends on the dynamically evolving nonequilibrium
distribution of carriers.
Therefore, when evaluating the scattering probabilities associated with plasmas
far from equilibrium, it is not sufficient to assume equilibrium dielectric
functions, but instead one needs to evaluate the nonequilibrium dielectric
function of the plasma associated with the momentary carrier distribution.
For graphene this was, to the best of our knowledge, first pointed out in
Ref.~\cite{Tomadin2013}, where the dielectric function is evaluated for a two
component plasma in a quasi-equilibrium to describe the dynamic screening of
Auger processes, using a procedure described by Maldague \cite{Maldague1978,
VanDuppen2016}.
At the same level, the plasmon dispersion, defined by the zeros of the
dielectric function, $\varepsilon(q,\omega)=0$ is also functionally dependent on
the carrier distribution.
A recent experimental study has shown the plasmon response of graphene can be
activated by pulsed optical excitation, due to the increase in Drude weight
associated with generated non-equilibrium carriers \cite{Ni2016}.
It has been shown in Ref.~\cite{Chaves2016} that photoexcitation beyond the
Dirac cone can give support to plasmons with anisotropic dispersion relations.
In a previous work \cite{Page2015}, we have solved the plasmon dispersion for
photo-inverted graphene and found that it differs fundamentally from the
equilibrium dispersion, as the phase space for net stimulated emission grows and
plasmons can become amplified \cite{Hamm2015}.
However, the method used therein to calculate the plasmon dispersion is only
applicable to a two-component plasma in quasi-equilibrium and at temperatures
much smaller than the chemical potentials ($k_B T \ll \mu_e, \mu_h$).
Crucially, it does not accommodate for more general carrier occupations, such as
the presence of high-energetic carriers that are  generated during optical
excitation.

In this work, we show how to efficiently calculate the polarization function
(and hence the dielectric function), and plasmon dispersion of graphene for
arbitrary isotropic nonequilibrium carrier distributions within the
random phase approximation (RPA).
Whilst the formalism can be extended, we further assume the electronic dispersion
of graphene to be a gapless Dirac cone, i.e. as described by the massless Dirac
fermion (MDF) model.
Crucially, the method presented here fully accommodates for the
conditions encountered during pulsed photo-excitation: it is applicable to an
electron/hole plasma at high temperatures ($k_B T \gg \mu_e, \mu_h$) and allows
to incorporate the influence of high-energetic carriers.

This paper is structured as follows: firstly we present the general theory that
allows the polarization function for isotropic nonequilibrium carrier
distributions to be efficiently evaluated on the basis of  zero-temperature
equilibrium polarization functions.
Next, to determine the complex-frequency plasmon dispersion, we present a
contour-integration method that remains valid at high temperatures and when
taking into account the photo-generated high-energetic carriers characteristic
of a plasma far from equilibrium.
To demonstrate the versatility of the method we calculate the energy-loss
function and plasmon dispersion for three different cases: a high-temperature
equilibrium, a high-temperature inverted two-component quasi-equilibrium, and
a nonequilibrium case that combines a hot quasi-equilibrium bath with a
distribution of photo-excited high-energy carriers.

\section{Theory\label{theory}}
\subsection{Non-equilibrium polarization function \label{noneqPol}}
The Coulomb interaction of a pair of electrons in graphene is effectively
screened by the collective of electrons (the MDF plasma) according to
$V^{\mathrm{eff}}_\qv = V_q / \eps(\qv,\omega)$, where
$V_q = e^2/ ( 2\epsbar\eps_0 q )$ is the 2D Fourier transform of the Coulomb
potential, for excitations with frequency, $\omega$, and in-plane
wave-vector, $\qv$; and where $\eps(\qv,\omega)$ is the dynamic dielectric (or
screening) function. It can be expressed, within RPA,
as \cite{giuliani2005quantum}
\begin{equation}\label{epsRPA}
\eps(\qv, \omega) = 1 - V_q \Pi(\qv, \omega)
\;,
\end{equation}
introducing $\Pi(\qv, \omega)$, the polarization function, given by the bare
bubble diagram.
Importantly, the poles of the effective Coulomb potential,
$\eps(\qv,\omega)=0$, signify collective plasmon excitations, freely
propagating charge density waves that transport energy and momentum
\cite{Giuliani1982}.
It has been shown, that despite an interaction constant larger than one, the MDF
plasma of graphene behaves like a weakly interacting electron gas and
application of RPA is well-justified \cite{Hofmann2014}.

The polarization function which relates to the (non-local) sheet conductivity of
graphene, $\sigs(\qv, \omega) = i e^2 \omega / q^2 \, \Pi(\qv,\omega)$
\cite{giuliani2005quantum}, is calculated using the Lindhard equation
\cite{Wunsch2006,Hwang2007},
\begin{equation}\label{lindhard}
  \Pi[n](\qv,\omega) = \frac{g}{A} \sum_{\kv,s,s'}
  M^{ss'}_{\kv,\kv+\qv}
  \frac{
      n(\eeps^s_\kv) - n(\eeps^{s'}_{\kv+\qv})
  }{
    \eeps^s_\kv - \eeps^{s'}_{\kv+\qv} + \hbar(\omega + i\eta)
  }
  \;,
\end{equation}
in the limit as $\eta \rightarrow 0^+$.
The equation above introduces the in-plane electron wavevector $\kv$, indices
$s, s' = \pm$ labelling the conduction and valence band, electron degeneracy
$g$, and sheet area $A$.
Further, for graphene in the MDF approximation, the matrix element is
given by $M^{ss'}_{\kv,\kv'} = ( 1 + s s' \cos \theta_{\kv,\kv'} )/2$ and
electronic dispersion $\eeps^s_\kv = s \hbar \vf |\kv|$, with Fermi velocity
$\vf \approx c/300$.
In the scope of this work, we assume non-equilibrium carrier distributions
that are functions of energy,
$n(\eeps^s_\kv)$,
and that the carrier energy is isotropic in $\kv$ space, i.e.,
$\eeps^s_\kv = \eeps^s_k$.
It follows that the response will thereby be isotropic, so herein we drop the
vector character of $\qv$ in favour of $q$.

Fundamentally, the equation for the polarization and dielectric functions remain
valid in non-equilibrium as long as pair-excitations are not too highly damped,
so that it is possible to define non-equilibrium distribution functions,
$n(\eeps^s_k)$ \cite{Kadanoff1994}.
That is, if the relaxation rates of the distributions are long compared to the
oscillation timescales of the excitations, one can neglect non-Markovian
contributions \cite{Vasko2005}.
Experimental measurements of the ultrafast carrier dynamics of optically excited
carriers at 0.8~eV give a fastest relaxation timescale of 10-150~fs, followed by
slower cooling and recombination on a 150~fs-15~ps timescale
\cite{Dawlaty2008,George2008}.
At an oscillation period of 5.2~fs, such carriers would be well described by a
Markovian model, which permits the use of the model presented herein in cases of
equilibrium, two-component quasi-equilibrium, and for hot photoexcited
distributions.

A general expression for the non-equilibrium polarization function can be
constructed based on the equilibrium polarization (Eq.~\ref{lindhard}), as
first demonstrated in Ref.~\cite{Page2015}.
In the following we briefly present this result, before detailing how the result
is applied to both screening and plasmons.
Firstly, we make use of the delta function identity,
\begin{equation}
  \Pi [ n ] = \Pi \left[
    \int_{-\infty}^\infty \dd\eeps \:
      n(\eeps)
     \delta(\eeps-\circ)
  \right]
  \;,
\end{equation}
where `$\circ$' is the dummy variable which the outer $\Pi$ functional operates
over.
The $(q, \omega)$ dependence is omitted for clarity.
Then, as Eq.~\ref{lindhard} is a linear functional of the carrier distribution,
we are able to swap the order of the integration and the functional, so that
only the delta function remains inside the functional,
\begin{equation}
  \Pi [ n ] =
  \int_{-\infty}^\infty \dd\eeps \:
  \Pi \left[
     \delta(\eeps-\circ)
  \right]
  n(\eeps)
  \;.
\end{equation}
The delta function can be expressed as the derivative of a zero-temperature
Fermi-function, i.e. a Heaviside step function; the polarization function of
which is well known \cite{Wunsch2006,Hwang2007,Pyatkovskiy2009}.
This allows the polarization function to be expressed an integral transform of
the carrier density,
\begin{equation}
  \Pi [ n ] =
  \int_{-\infty}^\infty \dd\eeps
  \pd{\Pi_{\mu=\eeps}^{T=0}}{\eeps}
  n(\eeps)
  \;.
\end{equation}
The integral can be split into contributions from electrons and holes by
introducing carrier distribution functions, $\nE$ and $\nH$, where,
\begin{equation}
n(\eeps) = \nE(\eeps) \theta(\eeps) +
  \left[1 - \nH(-\eeps) \right] \theta(-\eeps)
\;,
\end{equation}
with the Heaviside step function, $\theta(\eeps)$, separating contributions
from the conduction and valence band.
The polarization function then takes form as a sum of the polarization
function of intrinsic graphene and contributions from electrons and holes,
\begin{equation}\label{elegant0}
  \Pi [ n ] =
  \Pi_{\mu=0}^{T=0} +
  \int_{0}^\infty \dd\eeps\,
  \big[
  \Pi'(\eeps) \nE(\eeps) + \Pi'(-\eeps) \nH(\eeps)
  \big]
\;,
\end{equation}
with $\Pi'(\eeps) = \pd{}{\eeps} \Pi_{\mu=\eeps}^{T=0}$ for brevity.
For systems with particle-hole symmetry, i.e. $\Pi'(\eeps) = \Pi'(-\eeps)$, such
as MDF graphene, Eq.~\ref{elegant0} can be written in terms of the joint
occupation $\nE + \nH$.
This can be further split into a sum,
$\nE + \nH = \sum_i n_i$, whose contributions to the polarization function can
be evaluated separately, due to linearity, i.e.,
\begin{equation}\label{elegant}
  \Pi [ n ] =
  \Pi_{\mu=0}^{T=0} +
  \sum_{i}
  \int_{0}^\infty \dd\eeps\,
  \Pi'(\eeps) n_i(\eeps)
\;,
\end{equation}
This is a general result that holds for any material with isotropic band
structure and carrier distributions, within the RPA, not just graphene.

For graphene in the MDF approximation, the intrinsic contribution is
\cite{Khveshchenko2006},
\begin{equation}\label{intrisicPol}
  \Pi_{\mu=0}^{T=0} =
  \frac{g}{8 \pi \hbar^2 \vf^2}
  \frac{
    -i \pi (u-v)^2
  }{
    4 \sqrt{u} \sqrt{v}
  }
  \;,
\end{equation}
where new coordinates
$u = \hbar(\omega + \vf q)/2$ and $v = \hbar(\omega - \vf q)/2$,
along the axes parallel and perpendicular to the Dirac cone have been
introduced.
These coordinates are particularly convenient to highlight the position of
branch points in the polarization function.
For graphene in the MDF model, the integral kernel is particle/hole symmetric,
$\Pi'(\eeps) = \Pi'(-\eeps)$, and takes the simple form
\cite{Page2015},
\begin{subequations}
\label{kern}
\begin{align} \label{kern.sum}
  \Pi'(\eeps) &=
  \frac{g}{8 \pi \hbar^2 \vf^2}
  \left(
    \tilde{\Pi}'_+(\eeps) + \tilde{\Pi}'_-(\eeps)
  \right)
\\ \label{kern.part}
  \tilde{\Pi}'_\pm(\eeps) &=
    \frac{
      2i\sqrt{\pm i(\eeps \mp u)} \sqrt{\pm i(\eeps \mp v)} \pm 2\eeps
    }{
      \sqrt{u} \sqrt{v}
    } - 2
\;,
\end{align}
\end{subequations}

\begin{figure}
 \includegraphics{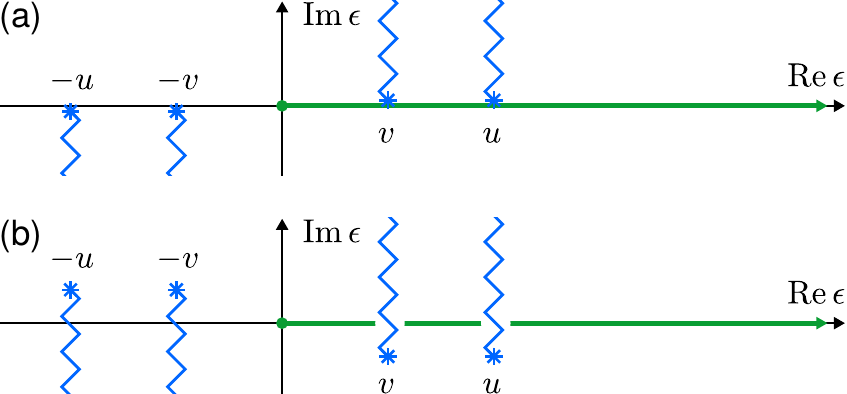}
 \caption{\label{contourReal}
  Complex-$\eeps$ plane diagram for the integration in Eq.~\ref{elegant}.
  The integration path is shown as the green line along the real axis.
  Branch points of the integration kernel (Eq.~\ref{kern}) are shown as blue
  `$\branch$' at $\eeps = \pm u, \pm v$.
  Branch cuts that stem from the branch points and are set vertically such that
  in (a) the function evaluates as if an infinitesimal positive imaginary
  part was added to $u$ and $v$, when $q$ and $\omega$ are real.
  In (b) $\omega$ is complex, so $u$ and $v$ have acquired finite imaginary
  parts, causing the branch points pierce the integration contour.
   }
\end{figure}
There are four branch points in Eq.~\ref{kern} coming from the square root
terms, $\eeps = \pm u$ and $\eeps = \pm v$, shown as blue `$\branch$' in the
complex-$\eeps$ plane in Fig.~\ref{contourReal}a, which lie on the real axis
for real values of $(q,\omega)$.
The branch cuts in  Eq.~\ref{kern} have been positioned vertically (see
Appendix~\ref{apndBranchCuts});
the integration can therefore be carried out directly on the real $\eeps$ axis
(green line), without the infinitesimal shift prescribed by Eq.~\ref{lindhard}.

Equation~\ref{elegant}, along with the definitions in Eqs.~\ref{intrisicPol} and
\ref{kern}, is all that is needed to evaluate the non-equilibrium polarization
function for real-valued arguments $(q,\omega)$.
As such it is sufficient when calculating the screening function and energy-loss
spectrum (see Sec.~\ref{eels}).
However, when tracing the plasmon pole $\epsrpa(q,\omega)=0$ for real-valued
wavevectors $q$, the plasmon frequency becomes complex \cite{Pyatkovskiy2009}
and the branch points in Fig.~\ref{contourReal} are no longer situated on the
real-$\eeps$ axis.
In this circumstance the branch points may cut through the integration contour,
as shown in Fig.~\ref{contourReal}b, and the value of the integral cannot be
uniquely determined.
The following section deals with the proper treatment of the integral in this
situation.

\subsection{Non-equilibrium plasmon dispersion relations \label{noneqPlas}}
Plasmons exist at frequencies where there is a pole in the effective Coulomb
potential, i.e. for the zeros of the dielectric function
(Eq.~\ref{epsRPA}), i.e.
\begin{equation} \label{epsRPAzero}
\epsrpa(q,\omega)=0
\;.
\end{equation}
The dielectric function is complex valued in regions where there is
Landau damping or electron-hole recombination, so it follows that for a real
wavevector, the roots themselves are at complex frequencies,
\begin{equation}
\omega = \omPl - i \gamPl
\;,
\end{equation}
where the real part is proportional to the energy of the plasmon, $\hbar\omPl$,
and the imaginary part is proportional to the net stimulated absorption rate of
the plasmon number density,
$2\gamPl = \gamma_\mathrm{stim} = \gamma_\mathrm{abs} - \gamma_\mathrm{emit}$.
Commonly, an approximation to the root is found by solving,
\begin{equation}
\re \epsrpa(q, \omPl) = 0
\;,
\end{equation}
for a real frequency, $\omPl(q)$, given a real $q$. The imaginary
part is then evaluated by taking a first-order Taylor expansion about $\omPl$,
perturbing by $-i \gamPl$, and solving, which leads to
\cite{Wunsch2006,Wang2007},
\begin{equation}
\gamPl =
\frac{\im \epsrpa(q, \omPl)}{\re \pd{\epsrpa}{\omega}(q, \omPl)}
\;.
\end{equation}
This approximation holds when the solved decay rate small with respect to the
frequency, $\gamPl \ll \omPl$, In this approximation however, as was shown in
\cite{Page2015}, the decay rate grows without bounds for increasing $q$, and
deviations from the exact complex solution are clearly visible in both real and
imaginary parts, as well as a failure to produce correct emission spectra in the
photo-inverted case.
This section shall detail how complex frequency plasmon dispersion relations can
be found for graphene with arbitrary carrier configurations.

Figure~\ref{contourReal} shows the integration of Eq.~\ref{elegant} in the case
of real valued frequencies;
there the integration contour is a straight line along the real axis from zero
to infinity.
When the frequencies become complex, the branch points detach from
the real axis and may cut through the integration contour, in which case the
polarization function evaluates incorrectly as the integral cannot be uniquely
defined.
Evaluation of Eq.~\ref{epsRPAzero} for complex frequencies thus requires a
deformation of the integration contour and the branch cuts such that both vary
continuously with increasing wavevector, $q$, and that neither cross each other
in complex-$\eeps$ space.

In general, an integral is independent of its integration contour;
any two paths with the same endpoints, that can be continuously deformed from
one to the other, will yield the same result, as long as no singularities are
crossed during the deformation.
Applying this principle to the polarization function, the contour of
integration, in Eq.~\ref{elegant}, can be altered by taking it off the real axis
to allow the branch points to move in a region below the real axis.

In addition to branch points, the other singularities that may appear in the
integrand are the poles of the carrier distribution function $n(\eeps)$.
The positions of these poles are fixed in the complex-$\eeps$ plane, independent
of the dynamic variables $(q,\omega)$.
As with branch points, the integration contour must not pass over poles
without being corrected for.
The Fermi function,
$n(\eeps) = (1+\exp((\eeps-\mu)/ T))^{-1}$,
for example, has its poles at the complex fermionic Matsubara frequencies,
$\eeps^{(n)} = \mu + i (2 n + 1) \pi T$ with residue $-T$.
If the integration contour passes over a pole, at $\eeps_\pole$,
the integral will gain a contribution due to the residue of the pole,
i.e., the integral of a closed contour encircling that pole,
$\mathcal{C}_\pole$.
For a pole in the carrier distribution function this contribution is,
\begin{equation}\label{residue}
\oint_{\mathcal{C}_\pole} \dd \eeps\: \Pi'(\eeps) n(\eeps) =
2 \pi i \Pi'(\eeps_\pole) \res_{\eeps_\pole}n(\eeps)
\;.
\end{equation}
Therefore, the integration contour in Eq.~\ref{elegant} can be altered to
pass over poles as long as the contributions of the residues are subtracted.

\begin{figure}
 \includegraphics{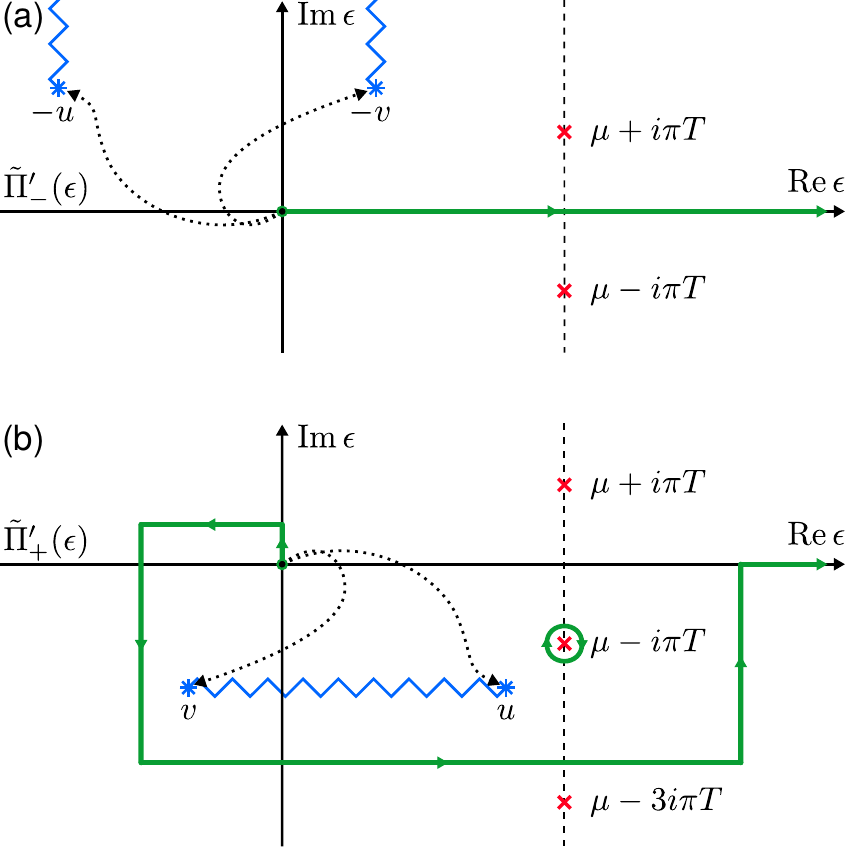}
 \caption{\label{contourDeformed}
  Complex contour integration for solving plasmon dispersion relations.
  (a) and (b) show the integration contour (green line) of
  $\int\dd\eeps \tilde{\Pi}'_\pm(\eeps) n(\eeps)$.
  Poles are shown as red `$\pole$', i.e. here for a Fermi distribution.
  Branch points are represented as blue `$\branch$'.
  The black dotted line illustrates how the branch points move as the solutions
  to $\eps(q,\omega) = 0$ evolve with increasing $q$.
  (a) integrates the $\tilde{\Pi}'_-$ kernel, over the positive real line,
  while (b) integrates $\tilde{\Pi}'_+$ over a contour that avoids being crossed
  by the moving branch points.
  One of the contained poles has been circled with a clockwise integration to
  indicate that the contribution of this pole must be subtracted, with the
  contour having passed over it.
  }
\end{figure}

Apart from the poles of the carrier distribution, which are explicitly
accounted for by Eq.~\ref{residue}, the prescription for calculation plasmon
dispersion curves is that the integration contour must not be crossed by
singularities as they move through $\eeps$ space, and that the contour is to be
deformed to enforce this.
It is useful here to break the integral into two parts, and calculate the
$\tilde{\Pi}'_+(\eeps)$ and $\tilde{\Pi}'_-(\eeps)$ contributions to the
integration kernel in Eq.~\ref{kern.sum} separately.
This will allow a different contour to be used in each part, and particularly
will separate the branch point singularities, so that only one pair has to be
considered in each integral.
For such evaluation, the favourable choice of branch cuts changes, as
illustrated in Fig.~\ref{contourDeformed} (see Appendix~\ref{apndBranchCuts}),
and Eq.~\ref{kern.part} is replaced by,
\begin{subequations}
\begin{align}
  \tilde{\Pi}'_+(\eeps) &=
    \frac{
      -2 \sqrt{\eeps - u} \sqrt{\eeps - v} + 2\eeps
    }{
      \sqrt{u} \sqrt{v}
    } - 2 \\
  \tilde{\Pi}'_-(\eeps) &=
    \frac{
      -2 i \sqrt{i(\eeps + u)} \sqrt{i(\eeps + v)} - 2\eeps
    }{
      \sqrt{u} \sqrt{v}
    } - 2
\;,
\end{align}
\end{subequations}
For the first equation, this keeps the branch cut compact, whereas the second
equation orients the cuts at $-u$ and $-v$ vertically upwards (blue zigzag
lines, as shown in Fig.~\ref{contourDeformed}).
The $\tilde{\Pi}'_-$ kernel is integrated along the real line but the
$\tilde{\Pi}'_+$ contribution follows a contour that, starting from zero, cups
the branch points $u$ and $v$ from underneath, and corrects for any poles that
would not have otherwise been included by a contour along the real axis. 

Given a form for the polarization function, it still remains to find a solution
to Eq.~\ref{epsRPAzero}.
For the complex frequency plasmon solutions, $\epsrpa(q, \omega)$ takes $q$ as a
real parameter, and a complex-$\omega$ is found for each $q$ which solves
Eq.~\ref{epsRPAzero}.
When treated as described herein, the dielectric function is well behaved as a
function of a complex-$\omega$ in the vicinity of solutions.
As such standard complex root finding methods, e.g. Newton's method,
that are available in numeric software packages, can be used.
The starting value for the root finding method, in the first instance, can be
the solution to an equivalent system with a Drude conductivity, as will be
described in Sec~\ref{sec:Drude} and subsequently a complex
frequency solution for a nearby $q$, that has been previously solved for,
should be used.

Examples of plasmon dispersion relations calculated in this way are presented in
the next section.

\section{High temperature equilibrium and quasi-equilibrium}
\subsection{Energy loss function and plasmon dispersion \label{eels}}

\begin{figure*}
 \includegraphics{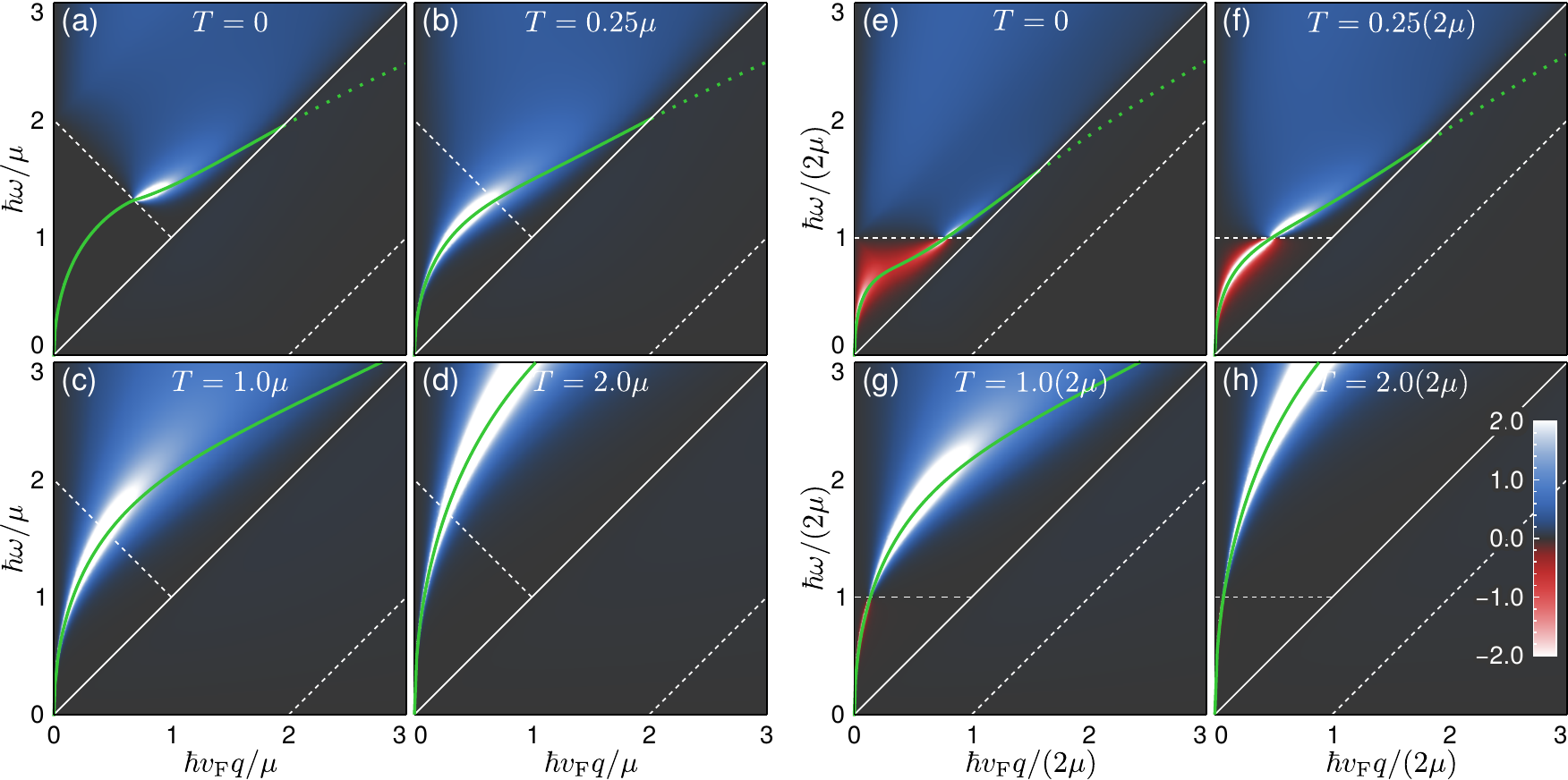}
 \caption{\label{finiteT}
  Energy loss function, $S = -\im \epsrpa(q,\omega)^{-1}$, over a range of
  temperatures.
  The solid white diagonal line is the Fermi velocity, whereas the dotted lines
  are the boundaries of regions of Landau damping at zero temperature.
  Complex frequency plasmon dispersion curves are plotted as green lines over
  each plot $(q, \re \omega_\mathrm{pl}(q))$.
  Parts (a-d): equilibrium graphene for temperatures in the range $[0,
  2\mu]$.
  Part (e-h): two-component plasma in quasi-equilibrium, following
  photo-excitation, with temperatures in the range $[0, 4\mu]$.
  The inversion line at $\hbar \omega = 2\mu$ separates regions of
  interband spontaneous emission (gain) and absorption (loss);
  in these regions, the loss function is negative and is shown in red.
  }
\end{figure*}

To demonstrate the contour integration method, we first evaluate the
complex frequency plasmon dispersion, $\omPl - i \gamPl$, together with the
energy loss function, $S = \-\im \epsrpa(q,\omega)^{-1}$ for both equilibrium
and two-component quasi-equilibrium, at finite temperatures, $T$ (given in
units of energy).
In both cases the carrier distribution functions for each band ($\nE$, $\nH$) are
Fermi functions, $n_{\mathrm{e}/\mathrm{h}} =
\{1+\exp[(\eeps-\mu_{\mathrm{e}/\mathrm{h}})/ T]\}^{-1}$.
The energy loss function is shown in Fig.~\ref{finiteT}, alongside
the corresponding complex frequency plasmon dispersion curves.
Parts (a-d) for equilibrium graphene, where there is a single Fermi-level,
i.e. the chemical potentials of each carrier species are of equal magnitude but
opposite sign ($\muE = -\muH = \mu$).
Part (a) is for zero temperature.
Here the Fermi edge is sharp and Landau damping is confined to regions permitted
by energy and momentum conservation.
The energy loss function, $S$, is finite in these regions and zero in the
so-called quasi-loss-free regions, the triangular regions in the plot.
The complex frequency plasmon dispersion (first solved at zero temperature
equilibrium in \cite{Pyatkovskiy2009}) is shown here as a green line.
The energy loss function is resonant about the plasmon dispersion curve, and is
broader where the plasmon loss $\gamPl$ is larger and sharper when it is small.
The solution of the plasmon dispersion curve continues over the Dirac cone
line, $\omega = \vf q$, into the intraband region. Notably this is not reflected
in the energy loss function as the plasmon curve has moved onto a different
branch of the energy loss function \cite{Kukhtaruk2015}.
Figure~\ref{finiteT}(b-d) show what happens to the plasmon dispersion and the
energy loss function as temperature is increased.
The energy loss function in this equilibrium case is in agreement with
\cite{Ramezanali2009}.
In the band, the Fermi edge stops being sharply defined and has a spread for
finite temperatures.
This results in the energy loss function not being confined to well defined
regions, becoming finite everywhere.
Again the energy loss function is resonant about the plasmon dispersion, and
this is more pronounced in regions of low loss, where the resonance is sharpest.
The plasmon dispersion curves themselves become steeper as $T$ is increased,
remaining in the interband region for a wider range of wavevectors.

Figure~\ref{finiteT}(e-h), accounts for the case of a
two-component quasi-equilibrium of excited intrinsic graphene.
In this case, each band has its own Fermi level, and here
($\muE = \muH = \mu$).
Parts (a-d) were scaled by the energy difference between the Dirac point and the
Fermi level, $\mu$, whereas parts (e-h) are scaled to the energy difference
between the two Fermi levels, $2\mu$, as was the convention in \cite{Page2015}.
With a population of electrons in the conduction band sitting above a population
of holes in the valence band, recombination of electron-hole pairs by stimulated
emission of plasmons becomes possible.
At zero temperature, as in Fig.~\ref{finiteT}(e), this manifests as the energy
loss function being negative, indicating gain (red on the plot), in the interband region for
energies less than $2\mu$.
For frequencies above this, stimulated emission is no longer possible, and
plasmons at these frequencies are absorbed by Landau damping (blue areas).
At finite temperatures, where the Fermi edge is not sharply defined,
phase space for stimulated emission at frequencies above $2\mu$ and for
absorption at frequencies below is allowed for.
The net rate of stimulated emission/absorption however remains sharply defined
at $\hbar \omega = 2\mu$, as reflected in parts (f-h).
This is also despite the plasmon curves steepening and growing as they did in
the equilibrium case, and hence dispersion curves for high temperatures having
gain in a comparably narrow wavevector range.

The manner in which the curves for equilibrium and two-component
quasi-equilibrium scale with temperature is similar in both cases.
In fact, for the highest temperature shown of each case,
Fig.~\ref{finiteT}, parts (d) and (h), the dispersion curves and loss functions
are almost identical.
At high temperatures, it is indeed the temperature that is responsible for the
scale of the features of the plasmon dispersion and energy loss function,
whereas at low temperatures, it is the chemical potentials that determine the
scale and features such as inversion.
In the next section, we shall introduce a scaling factor that generalizes for
arbitrary carrier distributions in order to more clearly show how the behaviour
of plasmons change depending on the carrier distribution.

\subsection{Drude weight scaling \label{sec:Drude}}

For regimes where temperature is much greater than the chemical potentials,
the distribution function and hence polarization function is, in relative terms,
insensitive to changes in either chemical potential.
This is the regime of thermoplasma polaritons, where the loss function and
plasmon dispersion scales linearly with temperature \cite{Vafek2006}.
In contrast, for low temperatures, the plasmon dispersion scales primarily
with the chemical potentials.

The Drude weight, defined as,
\begin{equation}
D = \lim_{\omega \rightarrow 0} \omega \im \sigs(0,\omega)
\;,
\end{equation}
can be used as a scaling parameter for the plasmons.
We introduce a derived frequency, $\omScale$, which later in this section we
shall plot plasmon dispersion curves in relation to:
\begin{equation}
\omScale = \frac{Z_0}{\alpha_0 g} D
\;,
\end{equation}
where $\alpha_0$ is the fine structure constant and $Z_0$ is the impedance of
free space.
To derive an expression for the Drude weight for a non-equilibrium
distribution in RPA, it suffices to use an expression for the intraband surface conductivity
in the local limit ($q \rightarrow 0$), given as \cite{Falkovsky2007b},
\begin{equation}
\sigma_\mathrm{intra}(0, \omega) = -\frac{i e^2 g}{4\pi \hbar^2 \omega}
\sum_{i}\int_0^\infty \dd \eeps\, \eeps \pd{n_i}{\eeps} \;,
\end{equation}
or more simply as,
\begin{align}
Z_0\sigma_\mathrm{intra}(0, \omega) = i \alpha_0 g \frac{\omScale}{\omega}
\;.
\end{align}
where the integral term is equal to the Drude weight frequency (after
integration by parts),
\begin{equation}\label{epsScale}
\epsScale = \sum_{i}\int_0^\infty \dd\eeps\:
n_i(\eeps)
\;,
\end{equation}
Substituting this into the solution for plasmons on a conducting sheet,
$1 + i c q Z_0\sigs(\omega) / (2 \epsbar \omega) = 0$,
returns the familiar square-root plasmon dispersion approximation,
\begin{equation}
\frac{\omega}{\omScale} = \sqrt{\frac{g\alpha_0}{2\epsbar}}
\sqrt{\frac{c q}{\omScale}}
\;,
\end{equation}
or even more compactly as, $\omtil = \sqrt{2 \alphag \qtil}$, for
$\omtil = \omega/\omScale$, $\qtil = \vf q / \omScale$, and
$\alphag = g \alpha_0 c/(4 \epsbar \vf)$.
which is the long wavelength limit for equilibrium distributions at zero
\cite{Wunsch2006,Hwang2007,Abedinpour2011} and finite \cite{Falkovsky2007a,
Ramezanali2009} temperature.
Thus, when the frequency and wavevector are scaled by the Drude weight, the
local limit of the plasmon dispersion is independent of the carrier
distribution.
We shall show later in this section that scaling by the Drude weight is indeed
useful to identify general behaviour when calculating the fully non-local
complex-frequency plasmon dispersion.

The Drude weight in Eq.~\ref{epsScale} is linear in carrier distribution, $n$,
and therefore can be calculated for each contribution independently and summed
for the total Drude weight.
For a Fermi distribution in a single band,
$n = \{1+\exp[(\eeps-\mu)/T]\}^{-1}$,
the contribution to the Drude weight takes the form,
\begin{equation}
\epsScale \rightarrow T \log\left(1 + e^{\mu/T}\right)
\;,
\end{equation}
which in the limit where chemical potential dominates, i.e.
$\mu \gg T$,
is purely
$\epsScale \rightarrow \mu$,
and in the temperature dominated regime,
$T \gg \mu$,
becomes
$\epsScale \rightarrow T \log 2$.

For equilibrium graphene, the electron and hole chemical potentials are equal
and opposite, $\mu_e = -\mu_h = \mu$.
The sum of both contributions to the Drude weight,
$\epsScale = T \log\left[\left(1 + e^{\mu_\elec/T}\right)\left(1 +
e^{\mu_\hole/T}\right)\right]$,
then becomes equivalent to the form,
$2 T \log \left[2 \cosh(\mu/2T) \right]$,
as appears in Ref.~\cite{Falkovsky2007a}.

\begin{figure}
 \includegraphics{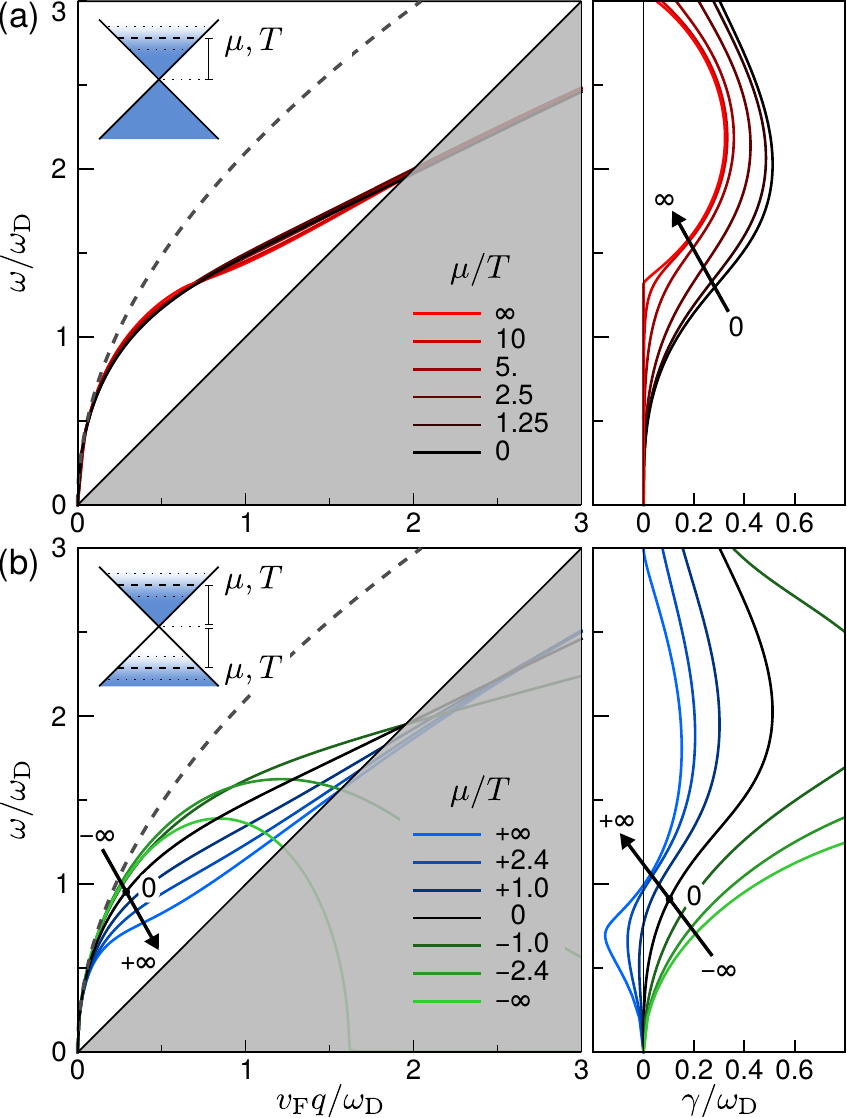}
 \caption{\label{scaleF}
  Plasmon dispersion curves at finite temperature for (a) equilibrium graphene
  and (b) two-component quasi-equilibrium.
  All curves are scaled to the Drude weight, $\omScale$, and
  plotted alongside the local Drude limit (dashed line).
  The curves progress from zero temperature (red in (a), blue in (b)), towards
  the case where temperature dominates over chemical potential (black curve,
  which is identical in (a) and (b)).
  For the two component case, (b), the curves continue for negative chemical
  potentials, i.e. for small probabilities of carriers in each band.
  Alongside the dispersion curves, scaled plasmon losses are plotted against
  frequency (right column).}
\end{figure}

Figure~\ref{scaleF} shows complex-frequency RPA non-local plasmon dispersion
curves scaled to the Drude weight $\omScale$.
This figure is for graphene in equilibrium in part (a) and two-component
quasi-equilibrium in part (b).
In each case the carrier distribution is a Fermi function in each
band (with both bands adding a contribution to the Drude weight) with a
temperature that is varied to span all ratios with the chemical potential.
Firstly, for equilibrium; part (a) shows that the scaled curves are largely
insensitive to the balance between chemical potential and temperature with all
curves overlapping.
For frequencies on the order of half the Drude weight or smaller ($\omega
\lesssim \omScale / 2$), the dispersion curves overlap with the square-root
Drude dispersion predicted by the local model.
Outside this frequency range, the curves each follow the zero-temperature
equilibrium solution (red curve).
Even in the high temperature limit, $T \gg \mu$, ($\mu/T=0$, black curve) the
dispersion is not significantly different from the zero temperature result.
Due to the similarity of these curves, they may all be approximated by a single
function, e.g.,
\begin{equation}
\frac{\omPl(q)}{\omScale} = \log\left(\frac{
  e^{
    \kappa(1 + e^{a \kappa})\sqrt{2 \alphag \qtil}
  } +
  e^{
    \kappa \left( a + b \qtil \right)
  }
}{
  1 + e^{a \kappa}
}\right)/\kappa
\;,
\end{equation}
which asymptotically approaches the Drude dispersion for $q \rightarrow 0$
and approaches a straight line at other values,
where $\kappa,a,b$ are parameters to be fit
(with values $\kappa=-2.79$, $a=1.06$, $b=0.45$ for suspended graphene).

The imaginary part of the complex frequency solution, i.e. the loss
curve, does vary with temperature.
This can be seen in the right panel of Fig.~\ref{scaleF}(a) where the red
curve, representing zero temperature, has frequencies where there is no loss,
i.e. where the plasmon dispersion passes through the loss free region.
When the temperature rises, absorption processes, that would be loss free at
zero temperature, are now no longer Pauli blocked and their
rates become finite, resulting in loss in the plasmon dispersion.
The extremal curve of the high temperature limit has a finite loss, and all
intermediate curves are bounded by the high and zero temperature limit.

In part (b) of Fig.~\ref{scaleF}, plasmon dispersion curves are drawn for
two-component quasi-equilibrium.
Here both positive and negative chemical potentials are solved for.
Whereas in the equilibrium case, with a single Fermi level, a positive or
negative chemical potential determines if the Fermi-level is found in the
conduction or valence band;
For a two-component quasi-equilibrium, each band has its own distribution
and Fermi level, and a positive chemical potential places the Fermi level within
it's band, and a negative outside of it.
Particularly a negative chemical potential indicates a low density of carriers
i.e. at the end of the Fermi distribution tail, where the occupation probability
for any state is less than one half.
Such distributions occur early in photo-excitation, when an excitation pulse
thermalizes and there are relatively few carriers but these carriers have a
large energy density.

In contrast to part (a), Fig.~\ref{scaleF}(b) shows that the plasmon dispersion
curves will spread out for different ratios of temperature to chemical
potential.
Each still follows the Drude square root behaviour for
$\omega \lesssim \omScale / 2$, but the curves fan out after this.
The black curve that represents temperature dominated behaviour
($\mu/T \rightarrow 0$)
is the same curve as in the equilibrium case, as in both cases the Fermi level
of each band is at the Dirac point.
From here, curves with a positive chemical potential fall underneath,
while curves with a negative chemical potential start to steepen over the
high-temperature curve. For sufficiently negative $\mu$, local extrema appear
in the dispersion, where the group velocity, $\pd{\omega}{q}$, is zero;
these are known as stopped-light points \cite{Tsakmakidis2014,Pickering2014}.
The extremal curve ($\mu / T \rightarrow -\infty$) has an
analytic solution, solved for in Appendix~\ref{lowCarriers}.
The loss curves show that for curves with a positive chemical
potential, there are frequencies ($\hbar \omega < 2\mu$) where the loss is
negative, i.e. there is gain.
As explained in the previous section, this is where the rate of
stimulated emission processes is greater than absorption processes.
In the high temperature limit and for negative chemical potentials, the bands
never have an occupancy probability greater than one half, resulting in greater
rates of plasmon absorption than emission at all frequencies.
For low carrier numbers ($\mu < 0$), the carriers in the system allow
plasmons to be supported, but these plasmons are far more likely to be absorbed
than to stimulate emission, and hence show the highest loss rates in the figure.

This section has shown that the shape of plasmon dispersion curves, once scaled
to the Drude weight, is approximately constant for equilibrium graphene, but
can vary quantitatively and qualitatively for non-equilibrium carrier
distributions such as a two-component quasi-equilibrium, especially for
configurations where there is a low density of high energy carriers in the
system.
Carrier inversion and high temperatures contribute to the gain and loss
channels of the plasmons such that the low-loss approximation,
$|\operatorname{Im} \omega| \ll \operatorname{Re} \omega$,
is no longer valid, and that plasmon dispersion and losses should be calculated
using this non-equilibrium complex-frequency procedure.
In the next section, we show how the non-equilibrium carrier distributions
produced immediately following photoexcitation can further affect the
polarization and plasmons.

\section{Photo-excitation non-equilibrium}

\begin{figure}
 \includegraphics{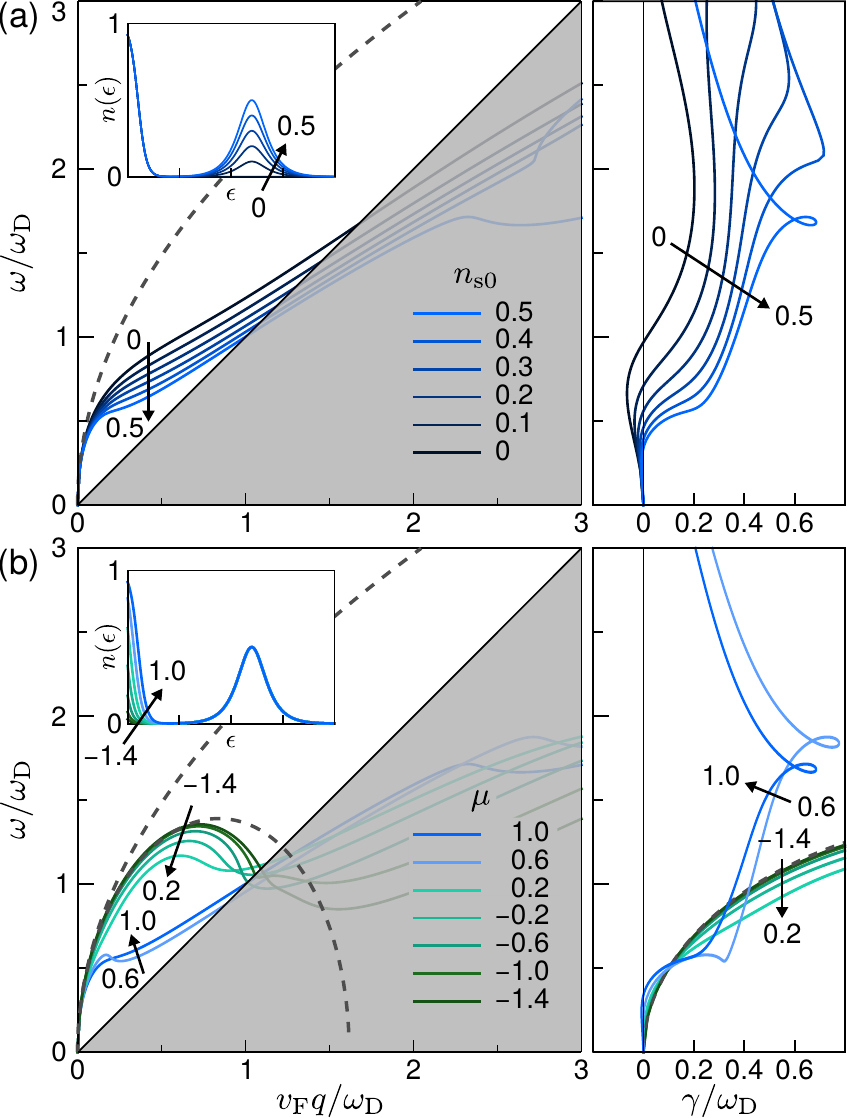}
 \caption{\label{photoex}
  Plasmon dispersion curves of photoexcited graphene with a carrier distribution
  function of a \emph{sech} excitation profile above thermalized Fermi bath,
  with parameters, $\muE = \muH$,
  $T=0.4\eV$,
  $\epsS = 1.2\eV$,
  $\gamS = 0.12\eV$.
  In part (a), the bath chemical potential is set at $\muE = 0.1\eV$ and the
  peak excitation occupation, $\nSo$, varies in the range $[0, 0.5]$ between no
  excitation and transparency.
  In part (b) $\muE$ varies in the range $[-1.4,1.0]$ with $\nSo$ fixed at 0.5.
  The carrier distribution function is inset in each part, and the plasmon loss
  plotted in the rightmost panels.
  All curves are scaled to the Drude weight, $\omScale$.
 }
\end{figure}

Until now, only polarization functions derived from Fermi distributions have
been shown.
The method presented in Sec.~\ref{theory} generalizes to calculations with
arbitrary carrier occupations.
We present a model for photo-excitation which describes a hot-carrier
nonequilibrium model.
That is, a transient population of photo-excited electrons are modelled to be
placed above a bath of thermalized quasiequilibrium carriers.

The photo-excited carriers are modelled with a hyperbolic secant (sech)
profile \cite{Chow2013}, above a background of relaxed quasi-equilibrium
carriers.
Both conduction and valence bands have the distribution,
\begin{equation} \label{photoexMod}
  \nEH(\eeps) = \frac{1}{1+\exp(\frac{\eeps-\mu}{T})} +
  \nSo
  \sech\left(\frac{\eeps-\epsS}{\gamS}\right)
  \;.
\end{equation}
The extra parameters introduced here are the excitation energy, $\epsS$
(which is half the pump photon energy);
the population width, $\gamS$
(which relates to the pump width and ratio of pump rate to electron
recombination rate);
and the peak occupation, $\nSo$, which ranges from 0, for no pump,
to 0.5 at transparency.
Calculating the complex contour integral of the sech contribution is almost
identical to that of the Fermi distribution;
like the Fermi distribution, the sech distribution has evenly spaced poles at
Matsubara frequencies,
$\eeps^{(n)} = \epsS + i (2 n + 1) \pi \gamS / 2$
with residue $-i (-1)^{n} \nSo \gamS / 2$,
therefore the same algorithm is employed.

Plasmon dispersion curves are shown for a photoexcited carrier distribution
function in Fig.~\ref{photoex}.
In part (a) the height of the sech peak is varied from 0.5 (i.e. transparency)
to 0 (no excitation).
The parameters, given in the figure caption, are such that when the sech peak is
at its maximum height ($\nSo = 0.5$), it contributes approximately twice as
much to the Drude weight as the Fermi bath does.
This contribution decreases linearly to zero as the height of the sech
peak is lowered.
The dispersion curves produced for each height of the sech distribution follow
the Drude dispersion in the local limit and then spread out for higher
wavevectors, being pushed down for as $\nSo$ increases.
This is in a similar manner to when temperature is increased in the
quasi-equilibrium case.
The loss of the plasmon dispersion curves increases with the height of the sech
peak, which is perhaps unexpected since in principal more carrier inversion is
being added to the system, however until these carriers relax within the band,
there are not enough emission channels at any particular energy to compete with
absorption and have a net gain for plasmons.
A more stark change comes in part (b) when $\nSo$ is held
constant at 0.5, and instead, the chemical potential of the thermalized bath is
reduced.
The curves split into two bundles, with those with the smallest chemical
potentials are attracted towards the low density-high energy limit, as in
quasiequilibrium for $\mu/T \rightarrow -\infty$ in Sec.~\ref{sec:Drude}
and Appendix~\ref{lowCarriers}.
In this case points of zero group velocity (turning points) appear in the
dispersion, and the dispersion is significantly different from that solved for
equilibrium graphene.

This section has shown that immediately after photo-excitation, whilst the
carriers relax initially to a two-component quasi-equilibrium, the momentary
plasmon dispersion and screening function can be qualitatively different.
Particularly, excited carriers need to relax within their band before they
contribute to plasmon gain processes.
The sech distribution has been used here to model photo-excitation, however a
sum of one or more sech functions may be used to fit other carrier
distributions, such as the non-thermal carriers observed during relaxation in
Ref.~\cite{Gierz2015}.

\section{Conclusion}
This work has outlined the procedure for efficiently evaluating RPA
polarization functions and plasmon dispersion relations in monolayer graphene
for arbitrary non-equilibrium carrier occupations.
This has allowed for these quantities to be calculated in cases of high
temperature, both in equilibrium and a two-component plasma quasi-equilibrium,
as well as in application to a model for carriers immediately following
photo-excitation, where they are momentarily excited to a ring of high
energy states.
The non-equilibrium carrier occupations explored in this paper are transient and
will evolve through a range of configurations as the system relaxes.
The irreducible polarization function, calculated here, is a key quantity of the
dynamic screening, which influences all carrier-carrier interactions, such as
Auger recombination and others which may play a role in carrier relaxation in
graphene \cite{Gierz2015,Giovanni2015}.
The theory is general and can be modified to describe graphene outside of the
Dirac cone regime \cite{Hill2009,Pyatkovskiy2009,Mihnev2015}, and indeed may be
applied to other two-dimensional materials \cite{Roldan2017} such as transition
metal dichalcogenides.

\section{Acknowledgements}
The authors acknowledge financial support provided by the Engineering and
Physical Sciences Research Council (United Kingdom).

\appendix
\section{Branch cuts \label{apndBranchCuts}}

\begin{figure}
 \includegraphics{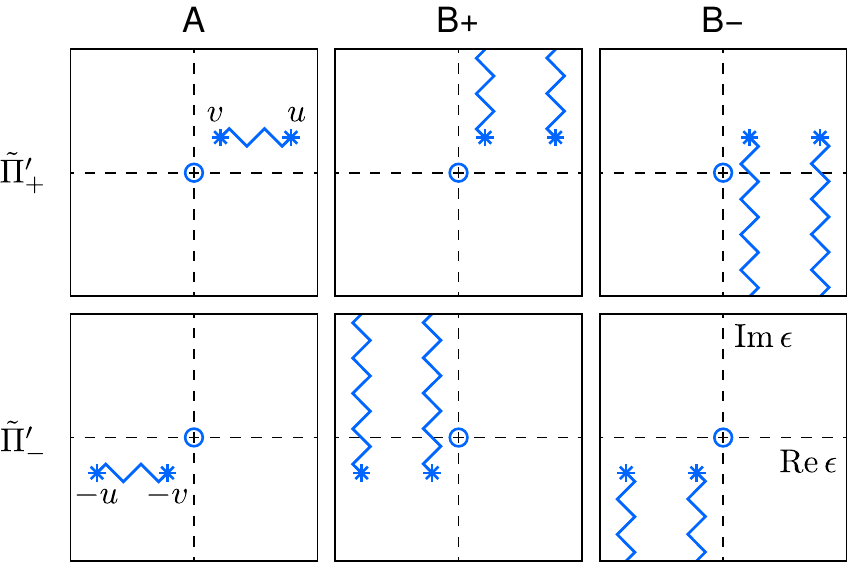}
 \caption{\label{branchcuts}
 Options for the position of branch cuts in the kernel function.
 The top row is for the $\tilde\Pi'_+$ function, which has branch points at
 $u$ and $v$, where the bottom row is for $\tilde\Pi'_-$ where the branch points
 are at $-u$ and $-v$.
 The columns represent choices for the branch cuts.
 \textsf{A} is a closed cut $[u,v]$ or $[-u,-v]$, where \textsf{B$\pm$} are open
 cuts such as $[u, \pm i \infty)$, etc, with their branch cuts starting from the
 branch point and extending to infinity.
 The circle at the origin of each diagram represents a zero of the function.
 On the principal branch of the kernels, $\tilde\Pi'_\pm(\eeps=0)=0$.
 }
\end{figure}

To manage the branches of the kernel functions (Eq.~\ref{kern.sum}),
a choice of branch cuts must be defined that anchor to the branch points.
Figure~\ref{branchcuts} shows a set of choices that have been used in this
paper.
Broadly, the choices available are closed or open branch cuts.
Closed cuts connect two branch points together and are finite in extent, whereas
open branch cuts start from one branch point and continue to infinity.
In the figure, these are labelled \textsf{A} for the closed cuts and
\textsf{B$\pm$} for open cuts with their endpoint at $\pm i \infty$.
In the paper, it was beneficial to use \textsf{B$+$} and \textsf{B$-$}
(for $\tilde\Pi'_+$ and $\tilde\Pi'_-$ respectively)
in Sec.~\ref{noneqPol} when evaluating the polarization function,
and to use \textsf{A} and \textsf{B$+$} in Sec.~\ref{noneqPlas} when solving
for plasmons.

Most computer implementations of the complex square root, $\sqrt{z}$, place the
branch cut on the negative real axis, choosing the branch with positive real
part.
The square root can be replaced with
$\sqrt{z} \rightarrow e^{i\theta/2}\sqrt{e^{-i\theta} z}$,
which rotates the branch cut counter-clockwise by $\theta$ from $0$ to $4\pi$,
i.e. two full turns.
The other branch can always be accessed by multiplying the square-root by $-1$,
i.e., $\sqrt{z} \rightarrow -\sqrt{z}$
This can be used to construct kernel functions with different branch cut
choices.

The \textsf{A} form is the simplest, and shall be used as a starting point,
\begin{subequations}
\begin{align}
  \mathsf{A}:\: \tilde{\Pi}'_+(\eeps) &=
    \frac{
      -2 \sqrt{\eeps - u} \sqrt{\eeps - v} + 2\eeps
    }{
      \sqrt{u} \sqrt{v}
    } - 2 \\
  \mathsf{A}:\: \tilde{\Pi}'_-(\eeps) &=
    \frac{
      -2 \sqrt{\eeps + u} \sqrt{\eeps + v} - 2\eeps
    }{
      \sqrt{u} \sqrt{v}
    } - 2
\;,
\end{align}
\end{subequations}
Individually, the square roots $\sqrt{\eeps-u}$, $\sqrt{\eeps-v}$ have a branch
point at $u$ or $v$ and their cuts extend to $\eeps \rightarrow
-\infty$.
In combination the two square root branches cancel as they overlap leaving a
branch cut between $u$ and $v$.
In order to calculate the \textsf{B} forms, these branch cuts are rotated by
$\pi/2$ clockwise for \textsf{B}$+$ and counter-clockwise for \textsf{B}$-$.
The principal branch is selected, by ensuring a root at zero for
$\im u = \im v > 0$.
Which yields the following as the \textsf{B} forms.
\begin{subequations}
\begin{align}
  \mathsf{B+}:\: \tilde{\Pi}'_+(\eeps) &=
    \frac{
      2i\sqrt{i(\eeps - u)} \sqrt{i(\eeps - v)} + 2\eeps
    }{
      \sqrt{u} \sqrt{v}
    } - 2
\\
  \mathsf{B+}:\: \tilde{\Pi}'_-(\eeps) &=
    \frac{
      -2i\sqrt{i(\eeps + u)} \sqrt{i(\eeps + v)} - 2\eeps
    }{
      \sqrt{u} \sqrt{v}
    } - 2
\\
\mathsf{B-}:\: \tilde{\Pi}'_+(\eeps) &=
    \frac{
      -2i\sqrt{- i(\eeps - u)} \sqrt{- i(\eeps - v)} + 2\eeps
    }{
      \sqrt{u} \sqrt{v}
    } - 2
\\
\mathsf{B-}:\: \tilde{\Pi}'_-(\eeps) &=
    \frac{
      2i\sqrt{- i(\eeps + u)} \sqrt{- i(\eeps + v)} - 2\eeps
    }{
      \sqrt{u} \sqrt{v}
    } - 2
\;.
\end{align}
\end{subequations}
As a final note, strictly speaking, the $\pm 2\eeps/\sqrt{u}\sqrt{v}$ terms are
superfluous as they cancel out.
They are included for numerical stability, such that each form limits to a
constant rather than growing linearly; this assists in the convergence of
numerical integrations.

\section{Highly energetic limit solution \label{lowCarriers}}

\begin{figure}
 \includegraphics{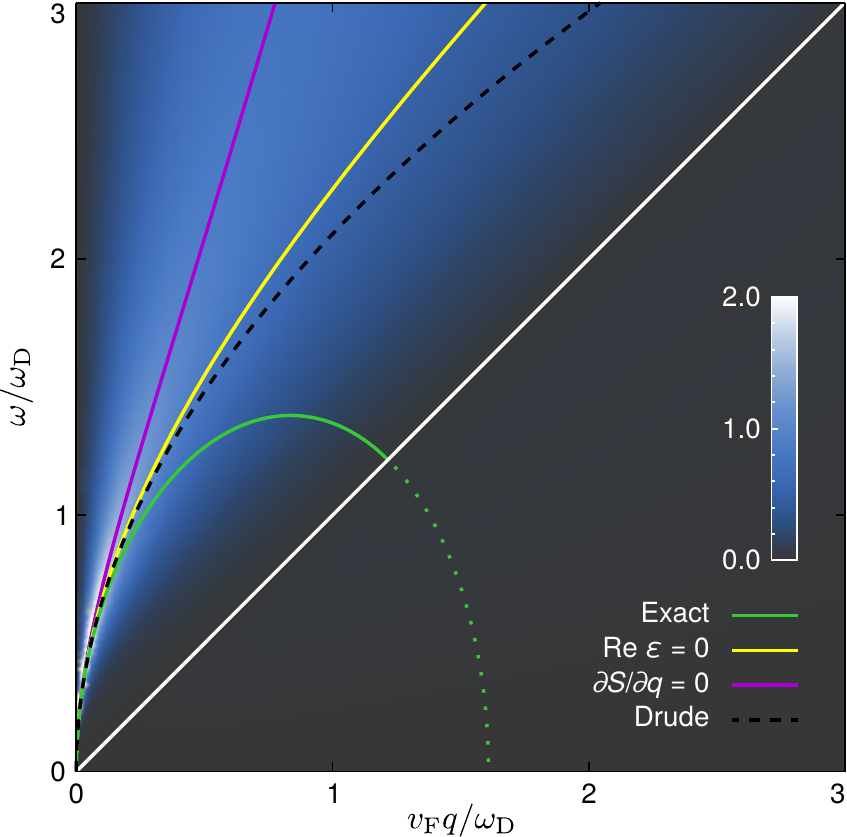}
 \caption{\label{exactPlasmon}
 Energy loss function, $S = -\im \eps(q, \omega)^{-1}$, for graphene in the
 highly energetic limit.
 Overlayed are the analytic plasmon solution (green), solved for in this
 section, alongside a number of approximations, i.e. the low loss approximation,
 $\re \eps = 0$ (yellow), following the peak of the loss function (magenta)
 and the local Drude model (black dashed).}
\end{figure}

In Figs.~\ref{scaleF} and \ref{photoex}, the plasmon dispersion curves that
corresponded to a small number of high energy carriers in the system were shown
to have turning points and curve downwards in an ellipse-like manner.
In the limiting case, this plasmon dispersion curve has an analytic solution,
shown in Fig.~\ref{exactPlasmon} which is worth examining because it is
demonstrably different to the approximations often used to describe plasmons in
the system, i.e. taking the low loss approximation that would solve
$\re\eps(q, \re\omega) = 0$ for a real frequency and taking a Taylor expansion
for the imaginary part, or alternatively by tracing where the peaks are in the
energy loss function.

To start we assume a delta-function occupation,
\begin{equation}
n(\eeps) = \hbar\omScale \delta(\eeps - \eeps_0)
\;.
\end{equation}
On face value, this would seem problematic since the occupation probability is
above one, however, this can be seen as an approximation function such as a
sech or a Gaussian with a sharp peak.
Putting this through Eq.~\ref{elegant}, and taking the limit, either as
$\omScale \rightarrow 0$ or equivalently $\eeps_0 \rightarrow \infty$,
i.e., $\epsScale \ll \eeps_0$, representing a small population of highly
energetic carriers; the dielectric function becomes,
\begin{equation}
\eps(q,\omega) = 1 + \frac{\alphag}{\qtil}\left(
\frac{i \pi \qtil^2 - 8\omtil}
{2\sqrt{\omtil+\qtil}\sqrt{\omtil-\qtil}}
+
4
\right)
\;,
\end{equation}
in terms of the scaled coordinates, $(\qtil, \omtil)$, introduced in
Sec.~\ref{sec:Drude}.
The complex zeros of this function can be solved for, returning the closed form
expression for the complex frequency plasmon dispersion,
\begin{equation}
\tilde\omega =
  \frac{(4\alphag + \qtil)\sqrt{
    8\alphag\qtil-((\frac{\pi\alphag}{2})^2-1)\qtil^2
  }}{8 \alphag + \qtil}
  -i \frac{2\pi\alphag^2\qtil}{8\alphag + \qtil}
\end{equation}
which reproduces the Drude limit,
$\omtil = \sqrt{2\alphag\qtil} - i\pi\alphag \qtil/4$,
as $\qtil \rightarrow 0$.

This curve, shown in green in Fig.~\ref{exactPlasmon}, is an attractor for
carrier systems that are dominated by high energy excitations, such as the two
component plasma when $-\mu \gg T$, or when carriers are photoexcited in an
otherwise empty band.

%

\end{document}